\documentclass[letterpaper, 10 pt, journal, twoside]{IEEEtran}

\usepackage{mathrsfs}
\usepackage{amssymb}

\usepackage{amsmath}
\usepackage{epsfig}
\usepackage{color}
\usepackage{algorithm}
\usepackage{algpseudocode}
\usepackage{hyperref}
\usepackage{xcolor}
\hypersetup{
  colorlinks   = true, 
  urlcolor     = black, 
  linkcolor    = black, 
  citecolor   = black 
}

\title{Unification of Consensus-Based Multi-Objective Optimization and Multi-Robot Path Planning}

\author{Michael P. Wozniak
\thanks{Purdue University}
}

\begin{document}

\maketitle

\begin{abstract}
Multi-agent systems seeking consensus may also have other objective functions to optimize, requiring the research of multi-objective optimization in consensus. Several recent publications have explored this domain using various methods such as weighted-sum optimization and penalization methods. This paper reviews the state of the art for consensus-based multi-objective optimization, poses a multi-agent lunar rover exploration problem seeking consensus and maximization of explored area, and achieves optimal edge weights and steering angles by applying SQP algorithms.
\end{abstract}

\section{Introduction and Motivation}
\subsection{Background}
\indent Lunar exploration is an increasingly relevant pursuit in the modern space era. The four phases of Space Development Theory (SDT) are exploration, expansion, exploitation, and exclusion \cite{Spacepower_Ascendant}. For private and government-backed space entities alike, all four phases of space development are intertwined with pursuing a long-term presence on the moon. Establishing this presence can enhance the United States' economic position by achieving a net-positive economic benefit from the resources offered by the Moon and beyond. \newline
\indent Several autonomy \& control challenges are associated with the establishment of an enduring presence on the moon. Autonomy is especially relevant because unmanned exploration offers increased efficiency, enabling cooperative completion of exploration without continuous human intervention. This importance is evidenced by NASA's pursuit of a cooperative trio of rovers that can cooperate without direct input from mission controllers \cite{NASA_JPL_Trio}. To this end, further research in autonomous algorithms for unmanned rovers would prove worthwhile for future exploration. \newline
\indent The assembly of a rover formation without continuous human input can be made possible by the alignment problem. This involves updating each agents heading by averaging its own heading and that of it's neighbors. While neighbors are considered as those within a specified radius, a small rover formation may consider all agents in the network to be neighbors \cite{AlignmentProblem}. Note that if one agent is non-cooperative, this agent's motion will guide the others until the network asymptotically aligns with the non-cooperative agent. It's important to note that this restricts the network to \textit{one} non-cooperative agent as the leader, and consensus/alignment cannot be reached if there are multiple non-cooperative agents. A research gap requiring further study is the case of the lead, non-cooperative rover under-performing. This may be caused by byzantine, or the lead agent sending conflicting information to its neighbors. This could also happen if the lead agent becomes malicious, pursuing goals that differ from the network's objectives. Another research gap is the idea of local minima being found via the optimization methods applied in this paper. As such, multi-start or other methods to support finding global extrema would be a helpful area for further research. Finally, a multi-objective optimization research gap that remains to be solidified is the selection of objective weights. While Pareto fronts have been developed in such problems \cite{CBMO}, the decision of objective function importance remains up to the end-users of these consensus applications.
\subsection{Literature Review}
\subsubsection{Consensus}
Consensus in multi-agent systems is a well-studied problem, and the fundamentals of consensus are employed in this paper. Recalling the system from \cite{AlignmentProblem}, consider a system of \(n\) autonomous agents  moving with the same speed but different headings. Each agent's heading \(\theta_i(t)\) can be updated using the average of its own heading and its neighbors at time \(t\). Each agent \(i\) has the neighbor set \(N_i(t)\), with edges \(\mathcal{E}(t)\). The described system can be modified to consider the \(n^{th}\) agent as non-cooperative, moving without considering input of its neighbors. Because this agent is non-cooperative, the other \(n-1\) agents will eventually converge to the heading of the non-cooperative agent such that it acts as a leader. \newline
\indent Mathematically, note that the update formula to average the headings of agent \(i\) and its neighbors at time \(t\) is posed as:
\begin{center}
\(\theta_i(t+1)=\frac{\theta_i(t)+\sum_{j\in N_i(t)}\theta_j(t)}{1+n_i(t)}\)
\end{center}
\indent \indent This update formula for each agent \(i\) will result in the system asymptotically approaching the same heading among all agents. This update formula is important to establish because it will be intertwined with the optimization being applied in this paper. Note that the preceding update formula assumes each agent is equally weighted. However, this may not always be the optimal update. To this end, let's generalize \(\theta\) to the state \(x\), and rewrite the averaging formula as:
\begin{center}
\(x_i(t+1)=\sum_{j\in N_i}w_{ij}x_j(t)\)
\end{center}
\indent \indent In this case, \(x_i(t)\) may represent heading, altitude, velocity, etc. Additionally, the case of different values among agents for \(w_{ij}\in [0,1]\) effectively introduces a weighted average. One method to select these values \(w_{ij}\) is Metropolis weighting. Considering the described multi-agent system with \(n\) agents, states \(x_i(t)\), the neighbor set \(N_i(t)\), and edges \(\mathcal{E}(t)\), the Metropolis weights as described in \cite{metropolis} can be assigned as:
\begin{center}  
\[w_{ij} =
\begin{cases}
1/(1+max\{d_i(t),d_j(t)\}) & \text{if $\{i,j\}\in \mathcal{E}(t)$} \\
1-\sum_{k\in N_i(t)} w_{ik}(t) & \text{if $i=j$} \\
0 & \text{otherwise}
\end{cases}
\]
\end{center}
\indent \indent This weight assignment is selected with consensus convergence and agreement as the priority.  The weights in the multi-objective optimization set forth in this paper will yield less agreement/consensus than Metropolis because they're fulfilling a second objective. \newline
\subsubsection{Multi-Objective Optimization and Consensus}\indent There are several recent studies that have delved into unifying multi-objective optimization and consensus problems. Such problems can be formulated considering weights \(\lambda_i\), and objective functions \(f_i(x)\), seeking to extremize \(f_{\lambda}(x)=\sum_{i=1}^p\lambda_if_i(x)\). The weights of each objective function can be tuned by varying the weight \(\lambda_i\).  Noting this framework, a 2022 study investigates a multi-swarm approach to approximating the Pareto front of multi-objective optimization \cite{CBMO}. This study compares static and dynamic weights, noting that dynamic weight adaption can couple the dynamics of different swarms to explicitly update the weights in each iteration. This paper also introduces a penalization strategy that avoids clustering on the Pareto front. This involves constructing a penalty term that is smaller when the objective is further away from the objective of the weighted means of the other swarms. In so doing, a spacing between node points on the Pareto front prevents clustering and enhances the approximation of the non-convex part of the Pareto front. \newline
\indent Another recent 2022 publication presents multi-objective consensus-based optimization methods via mean-field modelling. This optimization involves a set of interacting agents that explore the search space and attempt to solve all scalar sub-problems in parallel. The dynamics of these agents follow a mean-field model which facilitates algorithmic convergence. Based on \(N\) sub-problems generated for a multi-agent system, the limit of the step-size \(\Delta t\xrightarrow{}0\) and \(N\xrightarrow{}\infty\) otherwise known as the mean-field limit describes the agents dynamics. This approximation analytically describes the system behavior \cite{meanFieldApprox}.
\subsubsection{Rover Path Planning}\indent Rovers are continuously demanding more autonomy for increased efficiency \cite{frontiersREALMS} \cite{dynamicPathPlanning} \cite{autoRoversHumMars}. Additionally, multiple rovers being dedicated to a single mission is a common pursuit. For example, the Resilient Exploration and Lunar Mapping System (REALMS) investigates map coverage and system redundancy with two robots, noting the potential of scaling up to a larger swarm. REALMS invokes multiple robots to complete a single task to distribute mission risk and reduce mission costs \cite{frontiersREALMS}. \newline
\indent As part of NASA's Commercial Lunar Payload Services (CLPS) initiative, it pursues the demonstration of a network of multiple robots that can accomplish a single task autonomously. Multiple robots recording multiple measurements at the same time can record much more data than could be done by a single robot \cite{NASA_JPL_Trio}. \newline
\indent The preceding two contemporary examples of REALMS and CLPS demonstrate the immediate research need of increased autonomy and multi-agent applications in rover exploration.
\subsection{Further Research Contributions}
A key contribution of this paper is the application of Sequential Quadratic Programming (SQP) optimization to the problem of consensus. This is performed by considering the edge weights as design variables to be optimized. Additionally, this paper applies M.A.S. algorithms to the problem of maximizing explored area. While convergence speed is often the performance metric of interest, this paper poses a multi-objective optimization problem by considering convergence speed and explored area as two objective functions. \newline
\indent Note that the described optimization problem can be posed in several different ways. While it can be constrained for the rover formation to start and end in the same location, it can also be constrained to travel to a particular destination. To this end, there are several possible problem formulations that the proposed algorithm can solve. 
\subsection{Applications} The reviewed research areas have several possible end applications. While this paper considers a multi-robot system exploring the lunar surface, the unity of multi-objective optimization and consensus has many other possibilities. One example adjacent to the posed lunar exploration problem could be the collaboration between lunar rovers and lunar spacecraft. Because of the water ice on the lunar surface, hydrogen/oxygen propellant would be available in lunar missions to fuel spacecraft for aerial exploration \cite{Case_For_Space}. Similar to the multi-agent problem of surface-based rovers, this optimization problem could involve these spacecraft maximizing the number of rovers and/or ground stations that they transmit information to, yielding a strongly-connected system for lunar exploration. This concept is illustrated in the figure below. 
\begin{figure} [!h]
    \centering
    \includegraphics[width=0.75\linewidth]{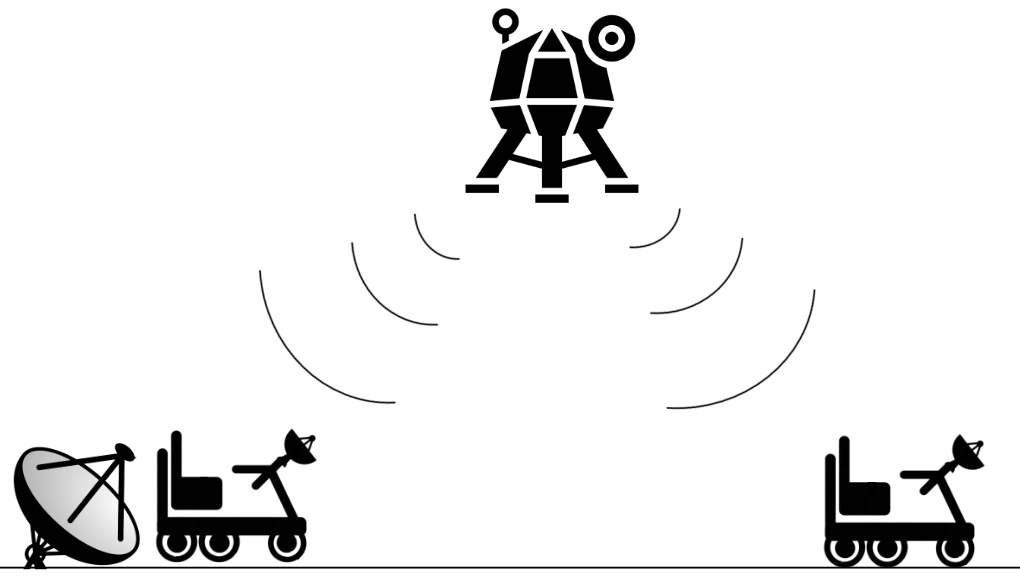}
    \caption{Multi-Agent Lunar Exploration via Spacecraft and Rovers }
    \label{fig:roverSC}
\end{figure}
\newline\indent There are other multi-objective consensus problems that are not  space-based. One example could involve applying consensus optimization to Collaborative Combat Aircraft (CCA) and minimizing fuel consumption, ultimately increasing mission efficiency. Another potential problem could be tuning the weights of an opinion dynamics network that not only secures rapid convergence to the same opinion, but also maximizes the efficiency of information transfer over the course of the interaction between agents.
\section{Problem Formulation}
\subsection{Multi-Agent System}
Consider a swarm of \(m\) rovers. There are \(m-1\) agents that are cooperative, with the \(m^{th}\) agent being non-cooperative. This implies that the \(m^{th}\) agent will guide the movement of the entire swarm, and that each agent \(i<m\) communicates with its neighbors \(N_i(t)\), which is considered constant in this context because the swarm will always be strongly connected and each agent will be in communicative range.  \newline
\indent Having established the (non-)cooperative nature of the system, it's important to note that the path of the non-cooperative rover is being directly optimized, driving the explored area optimization. The neighboring cooperative agents will abide by iterative update formulas that are a function of the weights \(w_{ij}\), which will be directly optimized to drive the consensus optimization. \newline
\indent The context of the proposed system is considered to be on the lunar surface, driven by the current plans of public and private space sectors, with more than 30 cislunar missions planned for execution by 2030 \cite{lunarExpPlans}. However, it's noted that this could be applied to any exploratory mission with multiple agents.
\subsection{Graph Representation}
\indent The graph representation of the \(m=4\) case is shown, noting bi-directional connections between agents \(1, 2\) and \(3\). Agent \(4\) provides information to all other \(m-1\) agents, but does not receive information from them.
\begin{figure} [!h]
    \centering
    \includegraphics[width=0.75\linewidth]{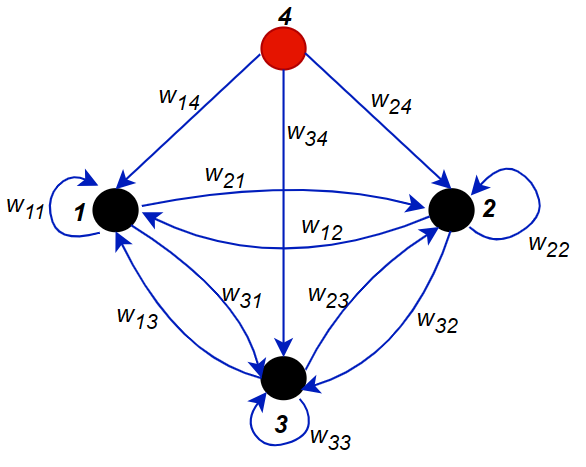}
    \caption{Graph Representation of Rover Swarm with \(m=4\) agents}
    \label{fig:graphrep}
\end{figure}
\subsection{Iterative Update Formulas}
\indent This paper considers the well-studied iterative update formula that results in all agents achieving the same value,  \(x_i(t+1)=\sum_{j\in N_i}w_{ij}x_j(t) \). Because the \(m^{th}\) agent is non-cooperative, or the \(4^{th}\) agent in the system being studied in this paper, applying this iterative update formula to the cooperative agents will result in all cooperative agents tending towards the same value as the non-cooperative agent. Note that \(x_i(t)\) will be considered to be the heading of the rover, such that the rovers will accomplish the same heading. \newline
\indent Weights are an imperative selection in the iterative update formula for the alignment problem set forth in this paper. Note that these weights are assigned with the objective of optimizing consensus convergence rate. Because this paper seeks to solve a multi-objective problem with explored area as another objective function, the weights will be determined by the results of SQP. Note that the heading of each agent \(i\) at time \(t\) is \(x_i(t)\). The weights will be subject to non-negative and row-stochastic constraints, namely \( w_{ij}>0:j\in N_i\) and \(\sum_{j\in N_i}w_{ij} = 1\), respectively. \newline
\indent Observe that the weights of the (non-)cooperative agents can be represented by the following matrix:\newline
\[ \bar{w}=
\begin{pmatrix}
w_{11} & w_{12} & w_{13} & w_{14}\\
w_{21} & w_{22} & w_{23} & w_{24}\\
w_{31} & w_{32} & w_{33} & w_{34}\\
w_{41} & w_{42} & w_{43} & w_{44}
\end{pmatrix}
\]
\indent \indent Noting that the \(m^{th}\) agent is non-cooperative and \(x_4\) headings are being optimized as design variables, the matrix form of iterative updates for cooperative agents simplifies to:\newline
\[
\begin{pmatrix}
x_{1}(t+1)\\
x_{2}(t+1)\\
x_{3}(t+1)
\end{pmatrix}=
\begin{pmatrix}
w_{11} & w_{12} & w_{13} & w_{14}\\
w_{21} & w_{22} & w_{23} & w_{24}\\
w_{31} & w_{32} & w_{33} & w_{34}
\end{pmatrix}
\begin{pmatrix}
x_{1}(t)\\
x_{2}(t)\\
x_{3}(t)\\
x_{4}(t)
\end{pmatrix}
\]
\indent \indent Noting these attributes of the edge weights, the optimization of these values can occur. While existing methods such as Metropolis weighting can ensure convergence to consensus, this paper will apply optimization tools to optimize the weights based on multiple objective functions.
\subsection{Optimization Problem Statement}
\indent This paper has recalled the (well-established) theoretical basis of consensus. Additionally, maximizing explored area is an integral part of mission success in space development. To unify these two phenomena, multi-objective optimization will be performed using SQP analysis and various test cases. The design variables will include the path of the lead, non-cooperative rover (enunciated in the form of heading at each iteration, assuming a constant step length for all rovers), and the non-zero edge weights in the network. \newline
\indent Let \(f_1\) represent the area explored, and \(f_2\) represent the consensus performance metric. Because this is a multi-objective optimization problem, consider the utopia points, \(f_{min,1}\) and \(f_{min,2}\), representing the respective objective functions when optimized independently. These will be employed in normalizing the objective functions when minimizing their linear combination. \newline
\indent The explored area \(f_1\) will be analyzed by a product of grid squares entered and maximum inscribed area. First, the explored surface can be gridded into \(C\) rows and \(C\) columns, where \(p\) is a counter variable for the row and \(q\) is a counter variable for the column. When one of the rovers in the swarm travels into the grid square \((p, q)\), this location will constitute as "explored" and a prescribed boolean variable \(\gamma_{pq}\) will be set to 1 (and set to 0, otherwise). Additionally, consider the maximum inscribed area to be \((max(X)-min(X))\times(max(Y)-min(Y))\), where \((X, Y)\) is the set of all traversed coordinates by all agents (not to be confused with \(x\), which represents heading). The maximum inscribed area is a relevant descriptor of exploration because a more diverse set of lunar samples would come from a path that reaches furthest in two orthogonal directions. This area term is quadratic which is necessary for the intended SQP algorithm. Having deduced the mathematics for grid squares entered and inscribed area, the first objective function \(f_1\) can be represented by their product:
\begin{center}
    \(f_1=(max(X)-min(X))\times(max(Y)-min(Y))\newline \times(\sum_{p=1}^{C}\sum_{q=1}^{C} \gamma_{pq})\)
\end{center}
\indent \indent Consensus will be quantified by the residual sum of squares (RSS). While the rate of convergence is usually employed to quantify consensus, such as in \cite{ConsensusCooperation}, it's noted that the problem in this paper has many possible solutions with a variable number of turns based on the explored solutions. This may serve as interference when the optimization algorithm is comparing various potential solutions, therefore the overall RSS error in heading among agents will be minimized to maximize convergence to consensus. First, the RSS error at each time \(t\) will be computed as \(RSS(t)=\sum_{i<m}(x_{i}(t)-{x_m}(t))^2\). This will effectively quantify the error between the heading of the cooperative agent \(i<m\) and the lead non-cooperative agent \(m\). Then, noting that this will be performed at each time-step, the consensus objective function \(f_2\) can be represented by summing the RSS terms over the entire timespan: 
\begin{center}
    \(f_2=\sum_{i<m}\sum_{t=t_0}^{T}(x_{i}(t)-x_m(t))^2\)
\end{center}
\indent \indent Observe that \(t\) is incremented by integers starting at 1, and \(T\) represents the total number of time-steps such that the cumulative heading error is being minimized. \newline
\indent Finally, constraints will be imposed onto the optimization problem. Note that inequality constraints will be harnessed for rover spacing, while equality constraints will be used for constraining final position and ensuring the weights sum to one. \newline
\indent Recall that the weights should sum to one such that \(\sum_{j\in N_i}w_{ij} = 1\). To this end, the equality constraints for each agent \(i\) can be posed as \(\sum_{j\in N_i}w_{ij}-1=0\). Additionally, the equality constraints for the final position can be specified within some user-defined tolerance. Noting that \(X_i\) is the set of all horizontal positions and \(Y_i\) is the set of all vertical positions for the agent \(i\), we can constrain the final position of agent \(i\) through the coordinate \((X_i(T), Y_i(T))\). As an equality constraint, this is formulated as \(X_i(T)-X_{target}=0\) and \(Y_i(T)-Y_{target}=0\), where the target final positions are application-specific. Because of the cooperative nature of agents \(i<m\), specifying the final position of agent \(m\) will guide the entire formation towards \((X_{target},Y_{target})\). \newline
\indent The inequality constraints can be imposed to ensure the desired spacing of the rovers. First, note that the maximum and minimum tolerances, \(maxTol\) and \(minTol\) respectively, represent the upper and lower bound for which the rovers should be spaced. To this end, the inequality constraint for each cooperative agent \(i\) can mathematically ensure minimum spacing via \(mintol-|X_i-X_m|<0\), and maximum spacing via \(|X_i-X_m|-maxTol<0\). This implies that the spacing between each agent will be between \(minTol\) and \(maxTol\). The same formulation is set forth for spacing in the \(Y\) direction. \newline
\indent Based on the preceding deductions, consider the formal optimization problem statement below: 
\begin{center}
\(\min{f = a_1\phi_1+a_2\phi_2}\)
\end{center} 
\indent \indent With the pseudo-objective functions:
\begin{center}
\(\phi_1 = \frac{f_{min,1}}{|f_1-f_{min,1}|},\phi_2 = \frac{|f_2-f_{min,2}|}{f_{min,2}} \)
\end{center} 
\indent \indent And the objective functions: 
\begin{center}
\(f_1=(max(X)-min(X))\times(max(Y)-min(Y)) \newline
\times(\sum_{p=1}^{C}\sum_{q=1}^{C} \gamma_{pq})\newline\newline
f_2=\sum_{i<m}\sum_{t=t_0}^{T}(x_{i}(t)-x_m(t))^2\)    
\end{center}
\indent \indent Subject to the equality constraints:
\begin{align*}
    \Bar{h}= \begin{pmatrix}
       w_{11}+w_{12}+w_{13}-1 \\
       w_{21}+w_{22}+w_{23}-1 \\
       w_{31}+w_{32}+w_{33}-1 \\
       w_{41}+w_{42}+w_{43}-1 \\
       X_4(T)-X_{target} \\
       Y_4(T)-Y_{target} 
     \end{pmatrix}
\end{align*}
\indent And the inequality constraints:
\begin{align*}
    \Bar{g}= \begin{pmatrix}
       minTol-|X_1-X_4| \\
       minTol-|X_2-X_4| \\
       minTol-|X_3-X_4| \\
       minTol-|Y_1-Y_4| \\
       minTol-|Y_2-Y_4| \\
       minTol-|Y_3-Y_4| \\
       |X_1-X_4|-maxTol \\
       |X_2-X_4|-maxTol \\
       |X_3-X_4|-maxTol \\
       |Y_1-Y_4|-maxTol \\
       |Y_2-Y_4|-maxTol \\
       |Y_3-Y_4|-maxTol \\
       X_4(T)-X_{target} \\
       Y_4(T)-Y_{target} 
     \end{pmatrix}
\end{align*}
\indent And the design variable vector \(\Bar{y}\), where
\begin{align*}
 &\Bar{y} = \begin{pmatrix}
       w_{11} \\
       w_{12} \\
       w_{13} \\
       w_{14} \\
       w_{21}\\
       \vdots \\
       w_{44} \\
       x(1) \\
       x(2) \\
       \vdots \\
       x(T)
     \end{pmatrix}
\end{align*}
\indent \indent Note that the objective function is a summation of the two pseudo-objective functions that are normalized by their respective utopia points. The first pseudo-objective function is a reciprocal of the second because of the maximization/minimization difference. While we're aiming to maximize the explored area which is enunciated as \(a_1\phi_1\), we're minimizing \(f\) and as a result, we must reciprocate the pseudo-objective function. Additionally, the weights of each objective function are denoted respectively as \(a_1\) and \(a_2\). This objective weight selection may be application-specific, but the combination should be convex such that \(a_1+a_2=1\).  \newline
\indent Observe that this optimization problem statement contains tunable, application-specific terms that are not design variables: \(X_{target}, Y_{target}, minTol,\) and \(maxTol\). Different target coordinates \((X_{target},Y_{target})\) will be set in simulations to allow the rover formation to be movable by algorithmic configuration. \(maxTol\) will be defined to keep the rovers in communicative range, and \(minTol\) will be defined to space the rovers enough for collision avoidance. \newline
\indent 
\section{Algorithm \& Main Results} \label{mainResults}
\subsection{Main Challenges}
There are several challenges associated with the proposed problem. The primary challenge is posing the optimization problem statement in a way that robustly achieves a solution optimizing both objective functions. This is especially prominent because of the multidisciplinary nature, in which utopia points must be identified and weights must be properly assigned to handle the division between optimizing consensus and explored area. Part of this challenge involves selecting a suitable quantity of design variables. While the SQP algorithm can find a minimum that is stationary within the step size tolerance, several local minima may be present in the usable/feasible space. The 
best way to handle this local minima challenge is through multi-start methods and diversifying the initial conditions across several runs of the algorithm. \newline
\indent Another challenge is that it becomes computationally expensive to optimize the turning angle at\textit{ each} step as a subset of the design variable vector. This complicates the process of searching for a minimum of the objective function, and increases the likelihood of finding one local minimum among many. Posing several design variables also complicates the process of scaling this problem up to many agents, which could be handled by grouping together equal weights when possible. One solution to this challenge involves posing a path plan that is periodic in nature, such that optimizing a subset of the problem can translate to optimizing the entire problem. Another solution can be assuming a symmetric weight matrix for the cooperative entries such that (\(w_{12}=w_{21}, w_{13}=w_{31}, w_{23}=w_{32}\)). This will reduce the number of design variables and enhance the computational efficiency of this algorithm. \newline
\indent An additional challenge is the intended optimization of explored area. The current method may not be the most robust measure of explored area for all applications. Alternate methods for the proposed \(f_1\) could be application-specific, such as quantifying explored area by assigning weights to various surrounding areas. For example, maximizing the diversity of lunar samples could involve collecting more samples from a particular unexplored area. This could also involve maximizing the time spent in sunlight if relying on solar power, or time near a shadowed crater if electrolysis units are stationed there.
\subsection{Key Idea}
The key idea is to implement the posed optimization problem statement into an SQP algorithm. In doing so, an optimal solution can be found that minimizes the difference in heading by achieving a consensus with tuned weights, and that maximizes the explored area. \newline
\indent Once the SQP algorithm has converged, the weights and turning angles will be stored in memory. Subsequently, the time-history of the optimal solution can be plotted for analysis, and the headings with time can be compared for various simulations. This approach delivers autonomy because the cooperative agents iteratively update their headings for alignment with the path of the lead agent, who observes pre-programmed optimal turning angles. The cooperative agents will continue to autonomously follow the lead regardless of their initial orientation or changes in direction from the lead agent. Imperfections in the lead agents execution of optimal turning angles will not be disruptive to formation control because the cooperative agents will iteratively update their headings based on sensor information.
\subsection{Algorithm}
The intended use of this algorithm is to perform an 
SQP optimization, then release the rover formation for optimal exploration. The cooperative agents will autonomously follow the lead agent, and feedback would decide if another optimization must be performed to further refine the edge weights and turning angles. \newline
\indent The provided pseudocode steps through the process of performing the SQP optimization \cite{SQP} then controlling the formation.
\begin{algorithm} [!h]
\caption{SQP Optimization \& Formation Control}
\begin{algorithmic}[1]
\State Define objective \(f\) and constraint functions \(\Bar{g}, \Bar{h}\)
\State Define distancing tolerances and target position
\State Define an initial guess $\Bar{y}$
\State Begin first iteration, $k \gets 0$
    \Repeat
        \State Solve Quadratic Programming subproblem
        \State Update $x$ using a step size $\alpha$
        \State Update the Hessian and gradient approximation
        \State $k \gets k + 1$
    \Until{convergence or max function evals exceeded}
\State \textbf{return} optimal solution, \(\Bar{y}^*\)
\State Command agent \(m\)'s optimal headings at each time-step 
\State Agents \(i<m\) steered autonomously per tuned weights
\end{algorithmic}
\end{algorithm}
\newline \indent Knowledge of the position of each rover is intertwined with the implementation of distancing tolerances \(maxTol,minTol\) and the final position (\(X_{target}, Y_{target}\)). For each agent \(i\) taking a step of length \(\alpha\) at each time-step, the position \((X_i,Y_i)\) is defined by:
\begin{center}
    \(X_i(t+1)=X_i(t)+\alpha*sin(x_i)\)
    \(Y_i(t+1)=Y_i(t)+\alpha*cos(x_i)\)
\end{center}
\indent \indent Note that a heading of \(0\)° is considered to be straight ahead, with positive turning angles being counterclockwise from the forward direction, and a step-length of 1 is considered. \newline
\indent Additionally, because this algorithm can be tailored to various applications, there are unique utopia points for each posed problem. To this end, in step 1 of the algorithm (define objective \(f\)), the algorithm must perform two preliminary SQP optimizations to find both utopia points when optimizing consensus and explored area independently. A less computationally expensive approach could also involve using Metropolis weights as the utopia point for consensus.
\subsection{Main Results}
\subsubsection{Optimization Convergence}
A key result of this paper is the robust convergence to optimal solutions. To ensure convergence, the initial guess \(\Bar{y}\) satisfied the majority of imposed constraints. This involved equal (row-stochastic) edge weights and rover spacing within the range \((minTol, maxTol)\). The initial guess didn't observe a final position at the target location, but rather consisted of no turns and a final position \(T\) steps away from the starting point. The turning angles could then be dithered to satisfy the final position constraints. \newline
\indent The optimization algorithm converged to solutions with a final position at the target location. The optimal weights also summed to one, as necessary for robust consensus implementation. Finally, the algorithm ensured collision avoidance such that the rovers never attempted to occupy the same space at the same time. The usable/feasible region was thoroughly explored in each optimization, as the function count regularly exceeded 3000, with over 50 iterations per minimization. \newline
\indent Optimization convergence consistently took less than 1 second from the initial script call to arriving at an optimal solution. This implies that the SQP algorithm can be re-run during deployed applications if feedback indicates a need for adjustments.
\subsubsection{Consensus}
Another key result of this paper is the agreement between agents. Similar to the results in \cite{ConsensusCooperation}, the proposed algorithm yields heading agreement between agents. Additionally, resembling the results of \cite{AlignmentProblem}, the formation is guided by a single non-cooperative agent in the proposed algorithms. The algorithm in this paper secures agreement similar to the results of literature surrounding consensus and alignment, while fulfilling another objective function. This has deeper implications as future research may unify consensus with any other desired objective.
\subsubsection{Explored Path}
Finally, a key result of this paper is the algorithms ability to maximize explored area. While several possible paths may be taken to arrive at the target location, the controlled formation is found to take the path that maximizes the diversity of samples. In the case of this algorithm, optimal exploration is associated with exploring the most grid squares, while traversing the greatest distance in two orthogonal directions. However, the deeper implication is that the consensus algorithm may concurrently explore a path that extremizes user-defined objectives such as spending more time near water ice at permanently shadowed craters.
\section{Simulations}
\begin{figure} [!h]
    \centering
    \includegraphics[width=0.5\linewidth]{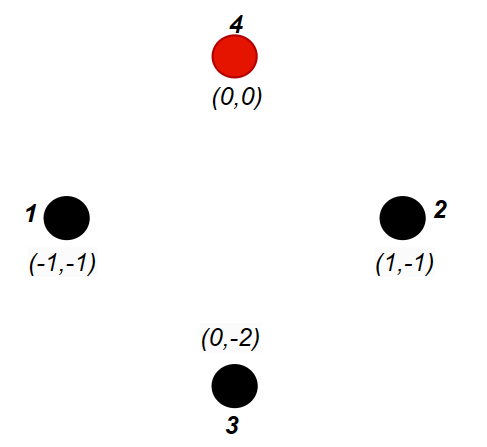}
    \caption{Initial Rover Positions}
    \label{fig:origin}
\end{figure}

Several simulations will be performed with the graph structure as shown in Figure \ref{fig:graphrep} (\(4\) agents) to illustrate the key conclusions of Section \ref{mainResults}. \newline
\indent Recall that the \(4^{th}\) agent is the non-cooperative agent, and all cooperative agents are iteratively updating heading to achieve the same value as their neighbors. Based on this configuration and the proposed optimization, an optimal solution \(\Bar{y}^*\) will be identified that minimizes the weighted sum of the two pseudo-objective functions. When an optimal solution is found, the optimal design values are noted and plotted for demonstration purposes. Additionally, the headings with time are noted for consensus analysis. \newline
\indent The initial positions are depicted below, noting a diamond formation with all agents facing forwards (the upward orientation of Figure \ref{fig:origin}).
\newline \indent The optimal edge weights will be presented with each simulation, however the turning angles will be omitted from the tabulated results as they are more intuitively presented in the figures of each traversed path.\newline
\indent Runs of the optimization algorithm consider an initial guess  assuming all edge weights are equal. Namely, the initial weight matrix is:
\[
\begin{pmatrix}
w_{11} & w_{12} & w_{13} & w_{14}\\
w_{21} & w_{22} & w_{23} & w_{24}\\
w_{31} & w_{32} & w_{33} & w_{34}
\end{pmatrix} =
\begin{pmatrix}
0.25 & 0.25 & 0.25 & 0.25\\
0.25 & 0.25 & 0.25 & 0.25\\
0.25 & 0.25 & 0.25 & 0.25
\end{pmatrix}
\]
\indent \indent Based on the preceding initial conditions, the algorithm will solve the Quadratic Programming problem until a stationary solution is identified.
\subsubsection{Simulation I}
\begin{figure} [!h]
    \centering
    \includegraphics[width=0.95\linewidth]{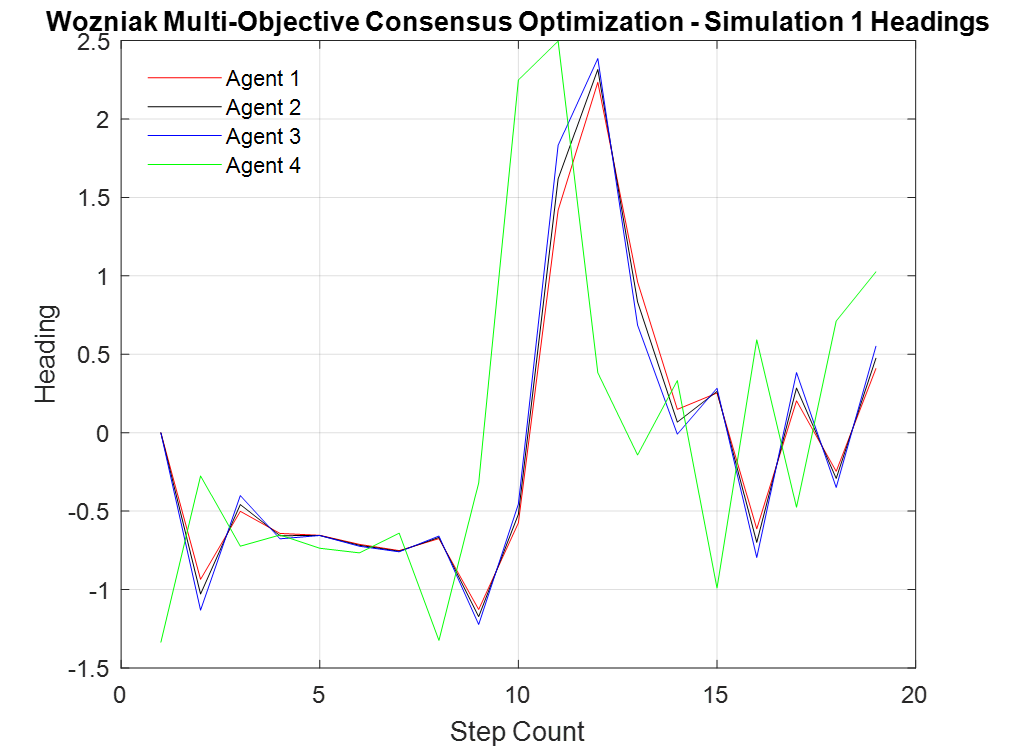}
    \caption{Simulation 1 -- Heading of each agent \(i\) with time}
    \label{fig:sim1_headings}
\end{figure}

\begin{figure} [!h]
    \centering
    \includegraphics[width=0.95\linewidth]{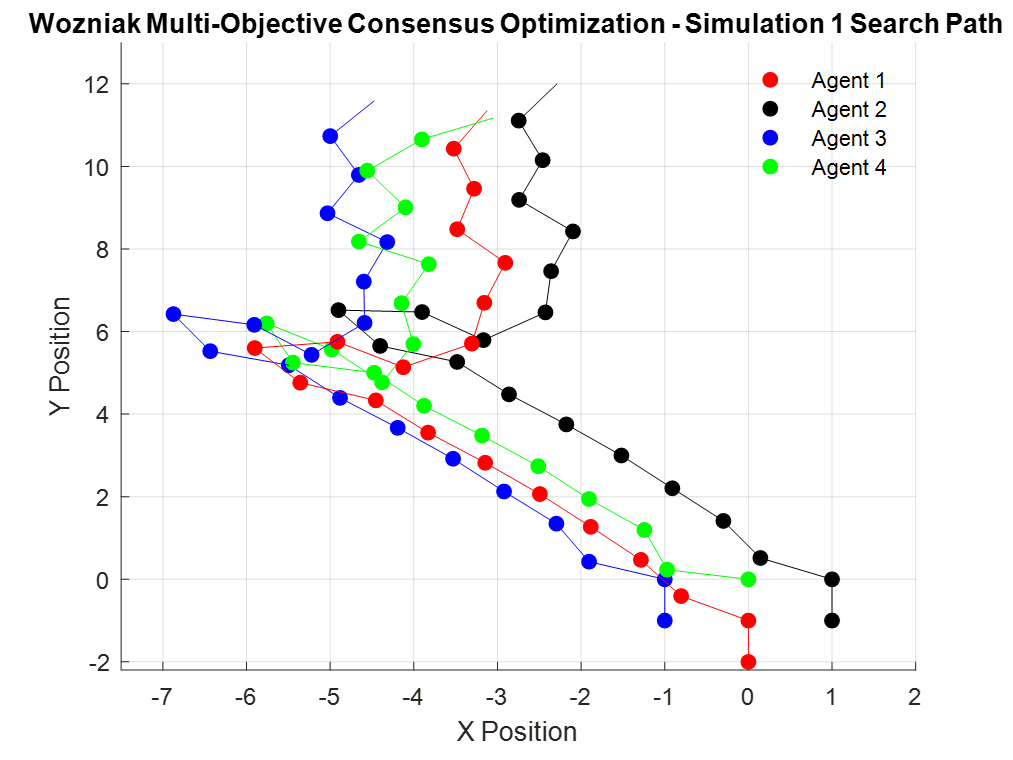}
    \caption{Simulation 1 -- Path Followed by Rover Swarm. For a video of this simulation, visit \url{https://youtu.be/YHBNw18-eTk}}
    \label{fig:sim1path}
\end{figure}

20-step, asymmetric edge weight matrix with equal weighting of pseudo-objectives (\(a_1=a_2=0.5\)), \(maxTol=5, minTol=0.2\), and a target position of \((X_4(T),Y_4(T))=(-3, 11)\) \newline
\indent \indent The initial guess of agent \(4\)'s path is \(x_4(t)\) = 0 for \(t\in[t_0,T]\) This case does not assume a symmetric weight matrix, such that \(w_{ij}\) does not necessarily equal \(w_{ji}\).  Upon running the SQP algorithm, the optimal edge weights are identified to be:
\[ \bar{w}^*=
\begin{pmatrix}
0.1000 & 0.1000 & 0.1000 & 0.7000\\
0.1000 & 0.1000 & 0.1000 & 0.7000\\
0.1000 & 0.1000 & 0.1000 & 0.7000
\end{pmatrix}
\]
\indent \indent Observe that the edge weights satisfy the imposed constraints, with each row being non-negative and summing to one. As such, the heading of each agent at time \(t+1\) is a convex combination of the agents headings at time \(t\). \newline
\indent The headings of each agent \(i<m\) with time \(t\) begin at the same value and largely coincide throughout the simulation. This is expected behavior of the implemented iterative update formula, and confirms that the consensus objective is being met. Note that the heading of the lead, agent \(4\), does not coincide with the headings of the cooperative agents because it does not receive information about the others. It solely guides the swarm while reporting its own state information and moving to maximize explored area.\newline
\indent Figure \ref{fig:sim1_headings} outlines the headings of each agent with time, noting that one step is taken at each \(t+1\). Observe that agent \(4\) guides the formation. While its heading does not coincide with agents \(1-3\), it serves a vital purpose because its heading changes direct the other agents. When the lead agent turns in a certain direction (appearing as up or down in Figure \ref{fig:sim1_headings}), the other agents turn towards that direction in the subsequent 1-2 steps. \newline
 \newline
\indent Based on the optimal headings/turning angles for the lead, non-cooperative agent, as well as the optimal edge weights, the path followed by the rover swarm is depicted in Figure \ref{fig:sim1path}.
 \newline
\indent Observe that the rovers followed a path with similar headings, following the same general direction. This solution is optimal for the proposed objectives.
\subsubsection{Simulation II}
\begin{figure} [!h]
    \centering
    \includegraphics[width=0.95\linewidth]{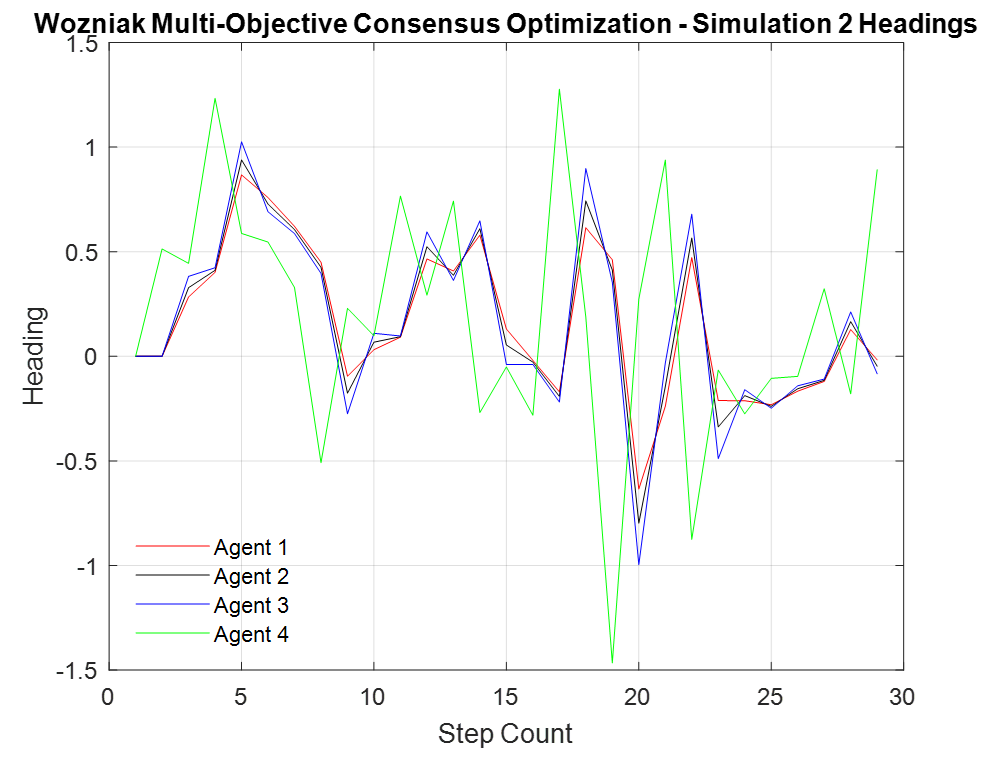}
    \caption{Simulation 2 -- Heading of each agent \(i\) with time}
    \label{fig:sim2headings}
\end{figure}

\begin{figure} [!h]
    \centering
    \includegraphics[width=0.95\linewidth]{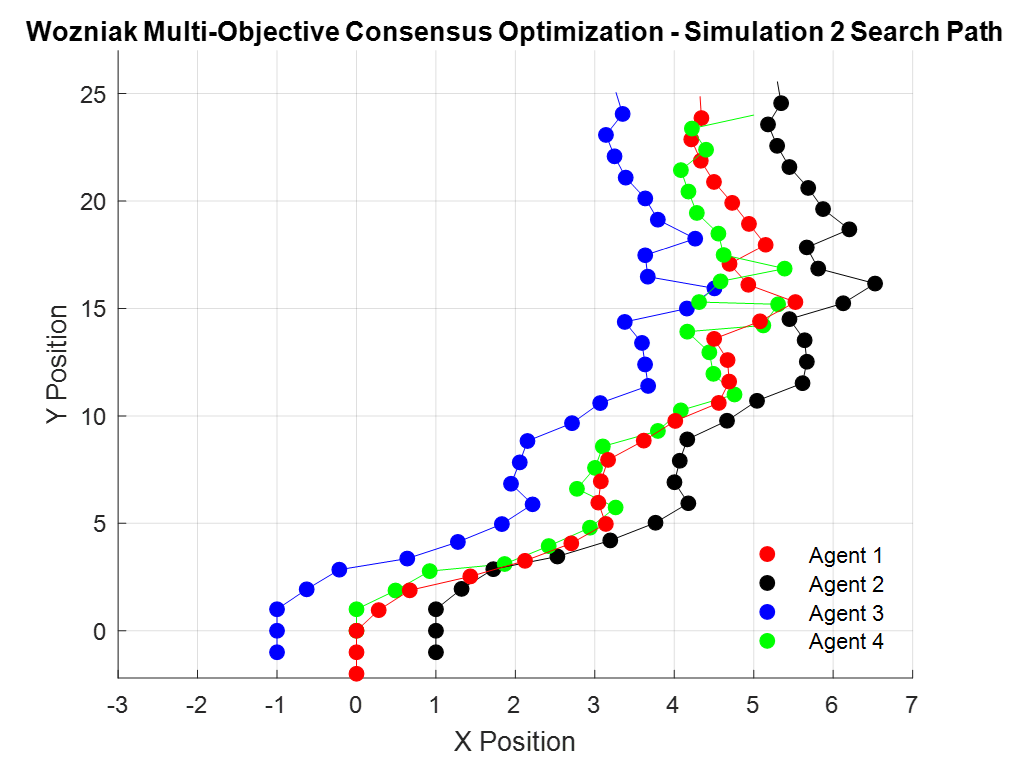}
    \caption{Simulation 2 -- Path Followed by Rover Swarm. For a video of this simulation, visit \url{https://youtu.be/54kTo6PeQMk}}
    \label{fig:sim2path}
\end{figure}

30-step, asymmetric edge weight matrix with equal weighting of pseudo-objectives (\(a_1=a_2=0.5\)), \(maxTol=5, minTol=0.1\), and a target position of \((X_4(T),Y_4(T))=(5, 24)\) \newline
\indent This simulation begins the SQP algorithm with the same initial conditions as described in Simulation I. It seeks to take 10 more steps than the preceding simulation and achieve a target final position further from the origin. In so doing, this simulation informs on the potential of scaling up. Nearly identical algorithm performance is observed when the number of steps is increased by a factor of \(1.5\), having achieved a feasible solution with tuned edge weights and turning angles. \newline
\indent The SQP algorithm being run for the conditions of Simulation II results in the following optimal edge weights that satisfy the constraints:
\[\bar{w}^*=
\begin{pmatrix}
0.1000 & 0.1000 & 0.2477 & 0.5523\\
0.3591 & 0.1000 & 0.1000 & 0.4409\\
0.2665 & 0.1000 & 0.1000 & 0.5335
\end{pmatrix}
\]
\indent \indent Based on the edge weights, the headings of each agent with time are depicted as seen in Figure \ref{fig:sim2headings}. Note that the case of more steps observes more turns, all guided by the non-cooperative agent. When agent \(4\) takes a sharp turn, the cooperative agents \(1-3\) turn the same direction after a phase shift.
\newline \indent Based on the headings shown, the path can be simulated as shown in Figure \ref{fig:sim2path}. Observe that the rovers tend towards the same direction and have achieved an optimal exploration of area subject to the imposed (in)equality constraints.
\subsubsection{Simulation III}
\begin{figure} [!h]
    \centering
    \includegraphics[width=0.95\linewidth]{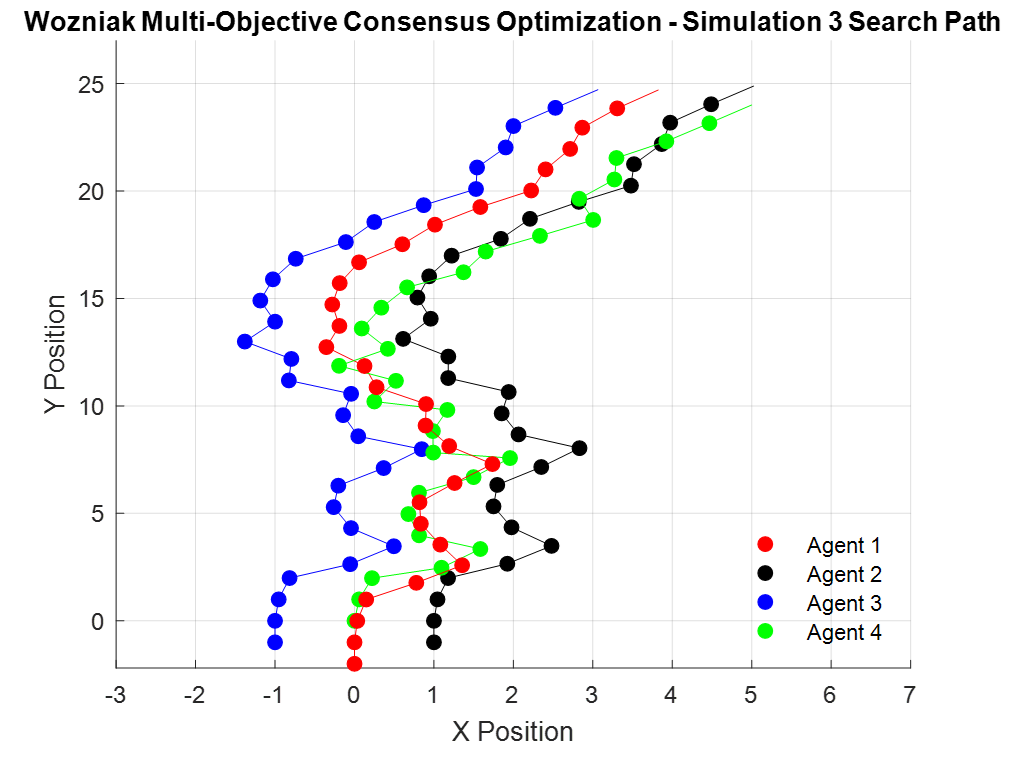}
    \caption{Simulation 3 -- Path Followed by Rover Swarm. For a video of this simulation, visit \url{https://youtu.be/_iAXiidjpYI}}
    \label{fig:sim3path}
\end{figure}

30-step, asymmetric edge weight matrix with unequal weighting of pseudo-objectives (\(a_1=0.25,a_2=0.75\)), \(maxTol=5, minTol=0.1\), and a target position of \((X_4(T),Y_4(T))=(5, 24)\) \newline
\indent The final simulation will illustrate an unequal weighting of pseudo-objective functions. In this case, the objective is 75\% weighted to consensus, and 25\% weighted to explored area. Convergence to an optimal solution is observed, satisfying the imposed constraints. \newline
\indent The SQP algorithm being run for the conditions of Simulation III results in the following optimal edge weights that satisfy non-negative and row-stochastic constraints:
\[ \bar{w}^*=
\begin{pmatrix}
0.2067 & 0.1000 & 0.1000 & 0.5933\\
0.1000 & 0.1000 & 0.1000 & 0.7000\\
0.1000 & 0.1000 & 0.1478 & 0.6522\\
\end{pmatrix}
\]
\indent \indent Based on the preceding optimal solution, the headings are plotted with time. Note that despite a greater weighting on the consensus objective, there is still a phase shift from heading changes by agent \(4\) and the turns performed by the cooperative agents. This demonstrates that tuning the objective weights will still result in a lag of cooperative agents because of the nature of the alignment problem. 
\begin{figure} [!h]
    \centering
    \includegraphics[width=0.95\linewidth]{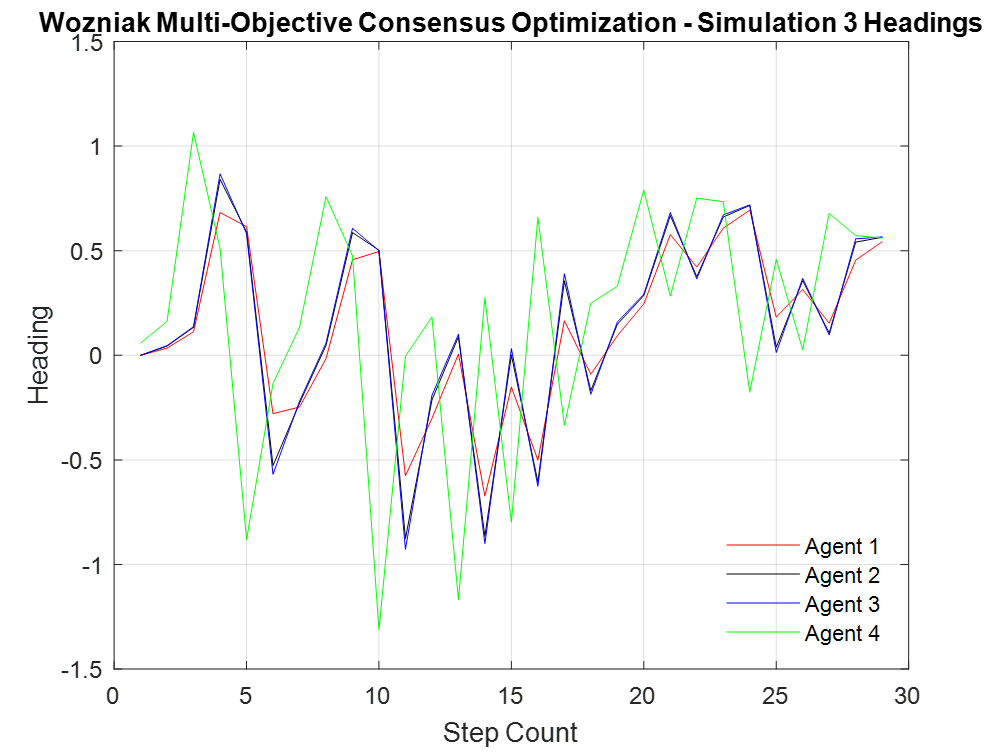}
    \caption{Simulation 3 -- Heading of each agent \(i\) with time}
    \label{fig:sim3headings}
\end{figure} \newline
\indent Based on the preceding headings, the time-history of the rover paths can be plotted as seen in Figure \ref{fig:sim3path}. Observe that the formation reaches the target location, and that it still maneuvers in the interest of increasing explored area.
\newline
\indent The successful optimization seen in these simulations proves that SQP optimization can enhance autonomy in space exploration. These algorithm runs could preliminarily be performed in deployed applications to determine the headings to command the lead agent, in which the cooperative agents would autonomously follow via received sensor information and tuned edge weights. Additional simulations of the presented algorithm configuration could introduce cases of symmetric edge weights, more agents, and a greater number of time-steps.
\section{Conclusion}
This paper has recalled existing literature surrounding three vital topics to autonomy \& control in space development: consensus, multi-objective optimization, and rover path planning. It discussed other use cases for the presented infrastructure, then formulated the multi-agent problem via graph representation and a formal constrained optimization problem statement. It proceeded to set forth and apply the SQP algorithm necessary to solve the presented optimization problem. Through multiple simulations, optimal solutions were depicted by plotting the time-history of the rover paths, as well as the headings of each agent with time. \newline
\indent While consensus-based multi-objective optimization has been robustly applied in this paper and several others, several research fronts remain for associated problems. \newline
\indent One potential extension to the algorithm provided in this paper is the unification of more elaborate path planning methods and consensus. This paper provided a low-fidelity model of path planning, noting a turning angle executed at each time-step \(t+1\) that subsequently influenced the amount of surface area explored. However, more elaborate path planning algorithms exist, including those that introduce multi-layer grids achieving objectives of collision avoidance and capturing terrain-vehicle interaction \cite{dynamicPathPlanning}. A more challenging but sophisticated approach to the posed multi-objective problem would be the implementation of consensus algorithms for a rover formation abiding by these kinds of dynamic path planning. \newline
\indent Another front for further research involves periodic grouping of path plans. While scaling up and introducing more time-steps may be a challenge because of the introduction of additional design variables, a mitigation method may involve posing the problem with a  periodic nature. For example, the formation may take one macro-step forwards at a time. Within each macro-step, it may follow an optimized path that maximizes explored area. Solving a periodic sub-problem and replicating it for several macro-steps would be less computationally expensive and better facilitate scaling up of more agents and more time-steps. \newline
\indent An additional area for further research could involve assigning weights to certain explored areas. For example, future lunar research missions may be in the presence of various ground stations that could include high-data-rate communication gear, solar panels, beamed microwave power, electrolysis units, and several other loads of equipment, as in the case of Moon Direct \cite{Case_For_Space}. This would motivate assigning higher value to lunar rovers spending time near these outposts based on the support available. Other examples include time spent near permanently shadowed craters, exploring where water ice may be present. Noting these motivations, further research into the explored area objective function could involve an unequal weighting of the landscape that overlays a mapping of where exploration is of higher value. This would be an invaluable way to maximize the resources spent on an exploratory missions such as the lunar problem set forth in this paper. \newline
\indent Finally, introducing a third dimension via teaming of land and air-based vehicles could be a highly valuable research pursuit. This need not be restricted to spaceflight applications and could involve Antarctic research \cite{antarctica}, natural disaster response \cite{naturalDisaster}, and many other possibilities. A relevant lunar-based extension to the algorithm set forth in this paper would be the teaming of lunar spacecraft and rovers. This may involve lunar landers on exploration-based flights that depart from and land on the lunar surface \cite{Case_For_Space}. The proposed problem could consider a model of where the ground stations and rovers are located, and maximize the quantity of downlinked information at each ground station, assigning higher weights to the most valuable ground stations.

\end{document}